\documentclass{PoS}

\usepackage{braket}
\usepackage{amsmath}
\usepackage{amssymb}
\usepackage{xspace}

\newcommand{\2}{\mathbf{2}}

\newcommand{\n}{\mathbf{n}}

\renewcommand{\k}{\mathbf{k}}

\renewcommand{\P}{\mathbf{P}}

\newcommand{\Fc}{\mathcal{F}}

\newcommand{\Jc}{\mathcal{J}}

\newcommand{\Mc}{\mathcal{M}}
\newcommand{\Oc}{\mathcal{O}}

\newcommand{\Rc}{\mathcal{R}}

\newcommand{\Wc}{\mathcal{W}}

\newcommand{\eg}{e.g.\xspace}

\newcommand{\ie}{i.e.\xspace}

\newcommand{\df}{\textrm{df}}
\newcommand{\nn}{\nonumber}
\newcommand{\diff}{\textrm{d}}

\newcommand{\beq}{\begin{equation}}
\newcommand{\eeq}{\end{equation}}

\title{Matrix elements of bound states in a finite volume}

\ShortTitle{Matrix elements of bound states in a finite volume}

\author{\speaker{Andrew W. Jackura}  \thanks{JLAB-THY-19-3060} \\
        Thomas Jefferson National Accelerator Facility, 
12000 Jefferson Avenue, Newport News, VA 23606, USA \\
Department of Physics, 
Old Dominion University, 
Norfolk, Virginia 23529, USA \\
        E-mail: \email{ajackura@odu.edu}}

\abstract{Recently, a framework was developed for studying form factors of two-body states probed with an external current. Finite volume matrix elements that may be computed via lattice QCD are converted to infinite volume generalized form factors. These generalized form factors allow us to study the structure of composite states. In this talk, we consider the application of this formalism to bound states, and compare the leading finite volume effects to the general results of the framework. Specifically, we consider the implications for the deuteron at the physical point, and conclude that it's necessary to use the full formalism to not be saturated by systematics}

\FullConference{37th International Symposium on Lattice Field Theory - Lattice2019\\
		16-22 June 2019\\
		Wuhan, China}

\begin{document}

\section{Introduction}

Mapping the excited hadron spectrum from QCD has become an established field in recent years. 
As most hadrons are resonances of multiparticle scattering channels, one must resort to techniques such as the L\"uscher's analysis, which connects infinite volume scattering amplitudes to finite volume spectra of hadrons in a box~\cite{Luscher:1986pf, Luscher:1990ux,Rummukainen:1995vs, Kim:2005gf, He:2005ey, Hansen:2012tf, Briceno:2014oea}. These analysis has proven effective in determining properties of many low-lying hadrons (see for example Ref.~\cite{Briceno:2017max} and references therein).

\begin{figure}[h!]
    \centering
    \includegraphics[ width=0.9\textwidth]{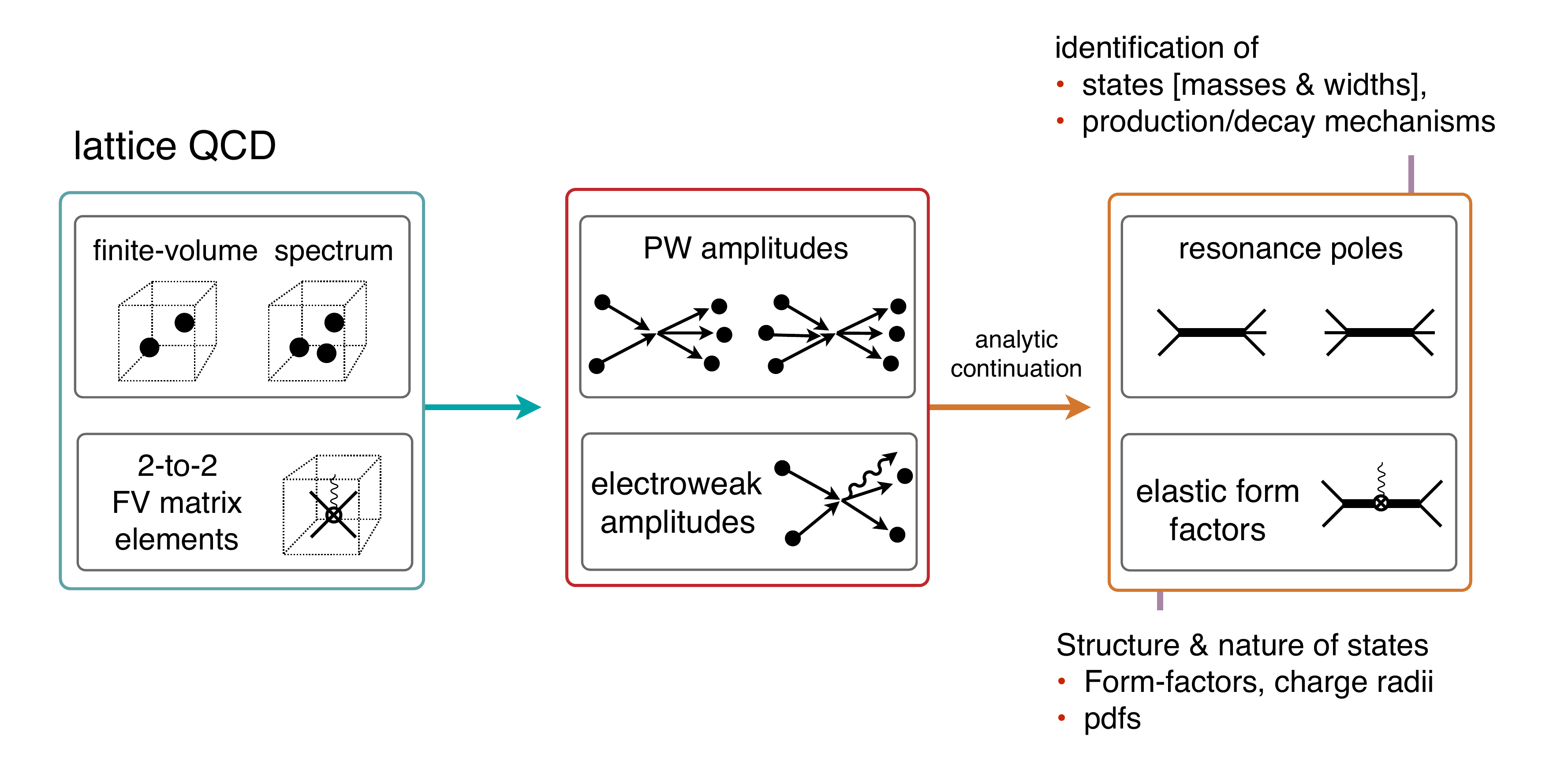}
    \caption{Roadmap for determining infinite volume scattering observables from lattice QCD calculations. Finite volume spectra and matrix elements, which are accessible through lattice QCD, are converted to infinite volume scattering amplitudes through relations such as the L\"uscher quantization condition and the presented $\2+\Jc\to\2$ formalism. From these amplitudes, one must analytically continue to the resonance or bound-state pole to determine properties such as mass, decay width, and form-factors.}
    \label{fig:road_map}
\end{figure}

Identifying the existence of an excited state, however, does not elucidate the underlying structure. The determination of form-factors, charge radii, parton distribution functions gives insight into the distribution of quarks and gluons of stable hadrons such as the proton. 
As in L\"uscher analyses, one can determine the structure of resonances and bound states within lattice QCD by relating finite volume matrix elements to these infinite volume observables.
Figure~\ref{fig:road_map} illustrates a road map for how one may attain, \eg form factors, of resonances and bound states.

To determine bound and resonant state form-factors, we consider a framework, first introduced in Refs.~\cite{Briceno:2015tza, Baroni:2018iau}, that allows one to determine infinite volume $\2+\Jc\to\2$ transition amplitudes via finite volume matrix elements. To gain confidence in this formalism developed, we provide two non-trivial checks on the formalism. First, we show that the charge associated with a conserved vector current is independent of finite volume corrections. The charge is protected from these corrections via the Ward-Takahashi identity, which relates the $\2+\Jc\to\2$ transition amplitude to the derivative of the $\2\to\2$ scattering amplitude. Second, we consider a scalar current to investigate the $L\to\infty$ limit of the scalar charge of a bound state. We refer the reader to Ref.~\cite{Briceno:2019nns} for detailed discussions of this study, where these results where first presented

\section{Two-hadron matrix element finite volume formalism}\label{sec:formalism}
We first review the finite volume formalism for two-hadron matrix elements. For simplicity, we will focus on the case of a vector current coupling to a system only in $S$-wave. Moreover, we consider a mass-degenerate two-particle system, which are distinguished by their charge: one which has charge $\textrm{Q}_0$ and the other neutral. The formalism can easily be be generalized for arbitary tensor currents and partial waves as in Refs.~\cite{Briceno:2015tza, Baroni:2018iau}. 

The formalism requires knowledge of the two particle scattering amplitude as well as the L\"uscher poles, which are the solutions of the L\"uscher determinant condition~\cite{Luscher:1986pf, Luscher:1990ux}. For systems in $S$-wave, the condition is a simple algebraic relation
\beq\label{eq:luscher}
\Mc^{-1}(s_n) = -F(P_n;L)
\eeq
where $s_n\equiv E_n^{\star\,2} = E_n^2 - \P^2$ is the L\"uscher pole, $P_n=(E_n,\P)$ is the four-momenta associated with the pole, $\Mc$ is the $\2\to\2$ $S$-wave amplitude, and $F$ is the finite volume function
\beq
F(P;L) = \left[ \, \frac{1}{L^{3}} \sum_{\k \in (2\pi/L) \mathbb{Z}^{3}} - \int \frac{\diff^3\k}{(2\pi)^{3}} \, \right] \frac{1}{2\omega_{\k}((P - k)-m^2 + i\epsilon)} \bigg|_{k^{0} = \omega_{\k}}
\eeq
with $\omega_{\k} = \sqrt{m^2 + \k^2}$. Equation~\eqref{eq:luscher} hold at the L\"uscher pole, \ie for the finite volume spectra of the two interacting particles in a box of size $L$.

Once the finite volume spectra and $\2\to\2$ amplitudes are known, we can turn our attention to $\2+\Jc\to\2$ processes. For the $S$-wave case, the formalism in Refs.~\cite{Briceno:2015tza, Baroni:2018iau} reduces to the relation
\beq\label{eq:BH1}
L^{3}\,\bra{P_{n}';L}\Jc^{\mu}\ket{P_{n};L} = \Wc_{L,\df}^{\mu}(P_{n}',P_{n}) \, \sqrt{ \Rc(P_{n}') \Rc(P_{n}) } 
\eeq
where $\Rc(P_n)$ is the generalized Lellouch-L\"uscher factor~\cite{Lellouch:2000pv,Briceno:2014uqa,Briceno:2015csa}
\beq\label{eq:LL}
\Rc(P_n) = \lim_{E \to E_n} \frac{E - E_n}{\Mc^{-1}(s_n) + F(P_n;L)},
\eeq
and $\Wc_{L,\df}^{\mu}$ is the finite volume quantity defined as
\beq\label{eq:BH2}
\Wc_{L,\df}^{\mu}(P',P) = \Wc_{\df}^{\mu}(P',P) + f(Q^2)\Mc(s') \, \left[ (P'+P)^{\mu}G(P',P;L) - 2G^{\mu}(P',P;L) \right] \, \Mc(s).
\eeq
The first quantity in Eq.~\eqref{eq:BH2} is the infinite volume $\2+\Jc\to\2$ transition amplitude, which can be written via an on-shell representation as
\beq\label{eq:Wdf_F}
\Wc_{\df}^{\mu} (P',P) = \Mc(s') \, \Fc^{\mu}(P',P) \, \Mc(s),
\eeq
where $\Fc^{\mu}$ is a generalized form-factor. In the kinematic region containing a bound state, $\Fc^{\mu}$ is a real function (see ~\cite{Briceno:2019nns,analytic}).

The functions $G$ and $G^{\mu}$ are finite volume functions which characterize power-law enhancement in the volume due to on-shell intermediate states of the triangle diagram. Explicitly, $G^{\mu}$ is defined as
\beq\label{eq:Gfcn}
G^{\mu}(P',P;L) = \left[ \, \frac{1}{L^{3}} \sum_{\k \in (2\pi/L) \mathbb{Z}^{3}} - \int \frac{\diff^3\k}{(2\pi)^{3}} \, \right] \frac{k^{\mu}}{2\omega_{\k}((P' - k)-m^2 + i\epsilon)((P - k)^2 - m^2 + i\epsilon)}\bigg|_{k^{0} = \omega_{\k}},
\eeq
with $G$ having an identical expression with exception that the $k^{\mu}$ in the numerator is absent. Features of this function can be found in Ref.~\cite{Baroni:2018iau}.

\section{Matrix elements of conserved vector currents}
We first investigate the $Q^2 = 0$ limit fo matrix elements of the temporal part of the conserved vector currents, \ie the finite volume corrections to the vector charge. Combining the equations in Section~\ref{sec:formalism}, we can write the finite volume matrix element as
\beq\label{eq:BH_vector}
 \bra{P_n;L} \hat{\textrm{Q}} \ket{P_n;L} = \frac{\Fc^{\mu = 0}(P_n,P_n) + \textrm{Q}_0 \left[ 2P_n^{\mu = 0}G(P_n,P_n;L) - 2G^{\mu = 0}(P_n,P_n;L)\right] }{ - \frac{\partial}{\partial E} \Mc^{-1}(s) \rvert_{E = E_n} + 2E_n G(P_n,P_n;L) - 2 G^{\mu = 0}(P_n,P_n;L)},
\eeq
where $\hat{\textrm{Q}} = L^{3}\Jc^{\mu = 0}$.
We see that the only difference in between the numerator and denominator is the first term.
The $G$ functions in the denominator arise from the derivative of the $F$ function.
An important property $\Wc_{\df}^{\mu}$ is found by considering the Ward-Takahashi identity,
\beq\label{eq:WTI}
\lim_{P'\to P} \Wc_{\df}^{\mu} (P',P) = \textrm{Q}_0 \frac{\partial}{\partial P_{\mu}} \Mc(s) = 2P^{\mu} \, \textrm{Q}_0 \frac{\partial}{\partial s}\Mc(s).
\eeq
Using Eq.~\eqref{eq:Wdf_F}, we can rewrite Eq.~\eqref{eq:WTI} as 
\beq\label{eq:WTI_genFF}
\Fc^{\mu = 0}(P,P) = - \frac{\partial}{\partial E} \Mc^{-1}(s).
\eeq
We notice that this exactly relates the first two terms of the numerator and denominator of Eq.~\eqref{eq:BH_vector}. We conclude that the the charge associated with the conserved vector current is independent of finite volume corrections.
This non-trivial result of the formalism provides our first consistency check against expectations of physical of the Ward-Takahashi identity. 
There is no expectation of the charge of an arbitrary tensor current to remain independent of finite volume corrections as the Lellouch-L\"uscher factor is independent of the Lorentz structure of the current.

\section{Two-hadron bound states and their matrix elements}
Since matrix elements of vector charge contain no finite corrections in the zero-momentum transfer limit, we consider now the bound state case of the system probed by a scalar current source $\Jc$. The particles are distinguished by the current, with one particle having a scalar charge $g_{S}$ while the other is neutral and we neglect any contributions from its composite nature. Still working with $S$-waves only, we consider a scalar bound state in the two-particle spectrum, \ie 
\beq
\Mc(s) \sim \frac{(ig)^2}{s - s_{\textrm{B}}} \qquad \textrm{as \,\,\,\, $s \to s_{\textrm{B}}$},
\eeq
where $s_{\textrm{B}}$ is the infinite volume bound state pole, with a mass $M_{\textrm{B}} = \sqrt{s_{\textrm{B}}}$, and $g$ is the bound state coupling. We can express the $\2+\Jc\to\2$ transition amplitude near the pole similarly
\beq
\Wc(s',s) \sim (ig) \frac{i}{ s' - s_{\textrm{B}} } F_{\textrm{B}}(Q^2) \frac{i}{s - s_{\textrm{B}}} (ig) \qquad \textrm{as \,\,\,\, $s,s' \to s_{\textrm{B}}$},
\eeq
where $F_{\textrm{B}}(Q^2)$ is the scalar form factor of the bound state and $F_{\textrm{B}}(0) \equiv g_{S,\textrm{B}}$ defines its scalar charge.
Alternatively, since the bound state is an asymptotic state, we may directly relate the scalar charge to the matrix element,
\beq
g_{S,\textrm{B}} = \bra{P_{\textrm{B}}} \Jc \ket{P_{\textrm{B}}}.
\eeq

Adapting the formulae laid out in Section~\ref{sec:formalism} for scalar currents ($G^{\mu} \to G$ and $\Wc^{\mu} \to \Wc$), we can investigate the finite volume corrections to the scalar charge. We define the finite volume scalar charge via the matrix element,
\begin{align}\label{eq:BH_scalar}
g^{\P}_{S,\text{B}}(L) & \equiv 2 E_B(L) L^3 \bra{P_{\text{B}},L} \Jc \ket{P_{\text{B}},L} ,
\end{align}
where the infinite volume bound state scalar charge is recovered in the $L\to \infty$ limit, \ie $g_{S,\text{B}}  \equiv \lim_{L \to \infty} g^\P_{S,\text{B}}(L)$. Using Eqs.~\eqref{eq:BH1} and \eqref{eq:BH2}, the finite volume scalar charge is
\begin{equation}\label{eq:fv_matrix_simple_scalar}
g^{\P}_{S,\text{B}}(L) =   \frac{  \mathcal F(s) + g_{S}  G(P,L)   }{ - \partial_s \Mc^{-1}(s)  +  G(P,L) - G^{ 0}  (P,L)/E  }  \Big\rvert_{P = P_{\text{B}}(L)} \,.
\end{equation}
Note we have used that the single particle scalar charge is $g_{S} = f(0)$. Equation~\eqref{eq:fv_matrix_simple_scalar} represents the all-orders expression for the finite volume scalar charge.

\begin{figure}[h!]
    \centering
    \includegraphics[ width=0.9\textwidth]{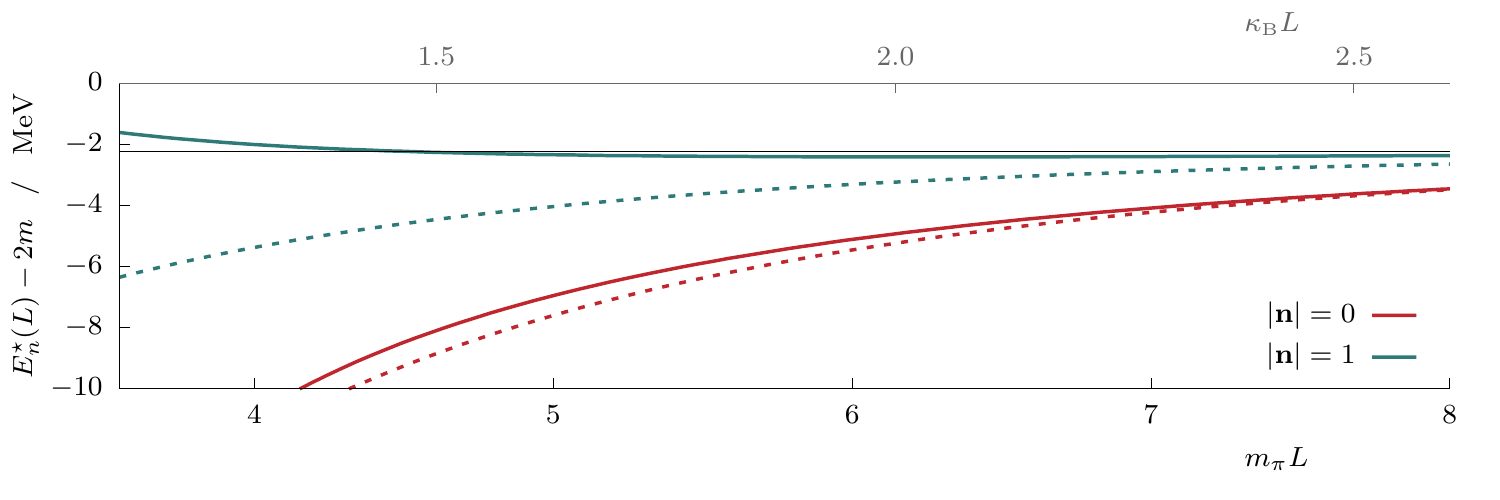}
    \put(-370,10){\colorbox{white}{(a)}}

    \includegraphics[ width=0.9\textwidth]{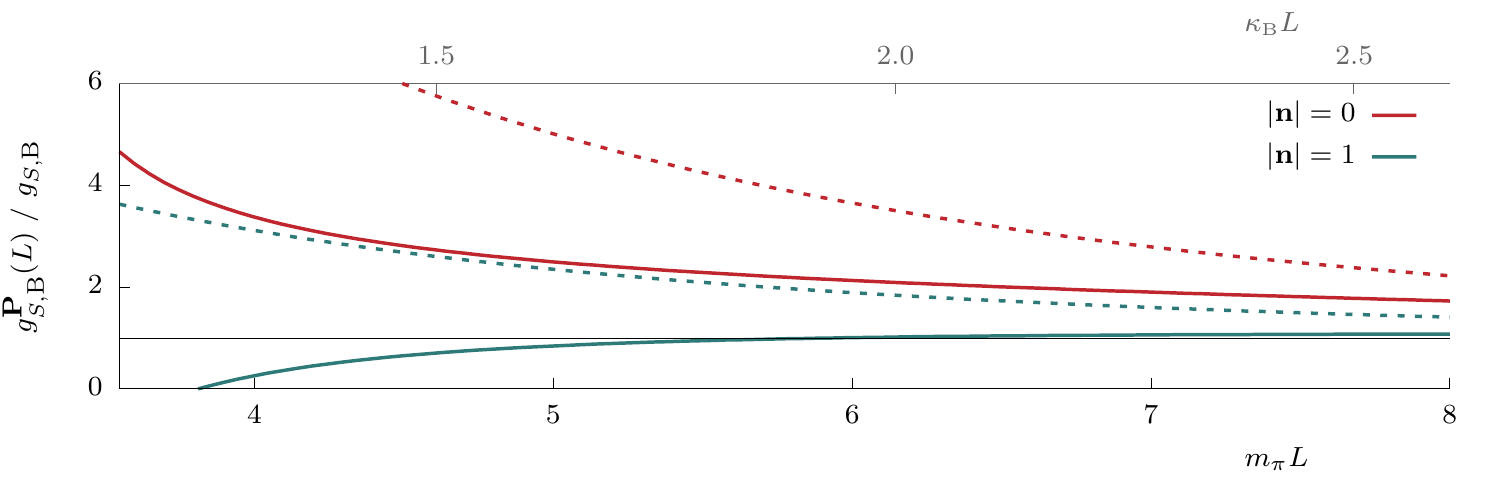}
    \put(-370,10){\colorbox{white}{(b)}}
    \caption{(a) Finite volume energy spectrum as a function of $m_{\pi}L$ for the $pn$-scattering parameters for a system at rest and boosted one unit in the $z$-direction with $\P = 2\pi \n / L$. The solid lines represent the exact spectrum form the L\"uscher condition and the dashed lines the spectrum at $\Oc(e^{-\kappa_{\textrm{B}} L} )$. The horizontal line at $E_n -2m \sim -2.21$ MeV represents the binding energy of the system with binding momentum $\kappa_{\textrm{B}} \sim 45.58$ MeV. (b) Ratio of the finite volume bound state form factor, $g_{S,\textrm{B}}^{\P}(L) = 2E_{\textrm{B}}(L) L^{3}\bra{P_{\textrm{B}},L}\Jc\ket{P_{\textrm{B}},L}$, to the infinite volume bound state scalar form factor $g_{S,\textrm{B}}$ at zero-momentum transfer as a function of $m_{\pi}L$. 
The energy spectrum extracted shown in (a) was used to evaluate the finite volume matrix element. The generalized form factor was assumed to be a simple energy-independent constant.  The horizontal line at $g_{S,\textrm{B}}^{\P}(L) \, / \, g_{S,\textrm{B}} = 1$ represents the deeply bound state result. Figure and caption taken from Ref.~\cite{Briceno:2019nns}.}
    \label{fig:fig-energies}
\end{figure}

It is now straightforward to expand Eq.~\eqref{eq:fv_matrix_simple_scalar} about $L \to \infty$. We first require the $L\to\infty$ limit for the bound-state L\"uscher pole,
\beq
s_{\textrm{B}}^{\P}(L) = s_{\textrm{B}} + \delta s_{\textrm{B}}^{\P}(L),
\eeq
where $\delta s_{\textrm{B}}^{\P}(L)$ is the finite volume correction to the infinite volume bound state pole $s_{\textrm{B}}$. One can show that for the pure $S$-wave case that Eq.~\eqref{eq:luscher} gives that the finite volume correction is of the form
\begin{align}  \label{eq:delta_sB}
\delta s^{\P}_{\textrm{B}}(L)  &=  g^2 F(P_{\textrm{B}},L)  + \Oc(e^{- 2 \kappa_{\text{B}} L})\, ,
\end{align}
where $F(P_{\textrm{B}},L)$ is the $F$-function evaluated at the infinite volume bound state pole, and $\kappa_{\textrm{B}} = \sqrt{ m^2 - s_{\textrm{B}} / 4 }$ is the binding momentum. 
Likewise, for $L\to \infty$, the matrix element Eq.~\eqref{eq:fv_matrix_simple_scalar} takes the form
\begin{align}\label{eq:BH_scalar_expand}
\frac{  g^{\P}_{S,\text{B}} (L)} {  g_{S,\text{B}}}
 & =   1 + \delta s^{\P}_{\text{B}}(L) \frac{\partial}{\partial s}  \bigg [  \frac{F_{\text{B}}(s)}{g_{S,\text{B}} }  + g^2  \frac{\partial}{\partial s} \Mc^{-1}(s)  \bigg ] \nn \\[5pt]
&  \qquad +  \frac{g^2 (  g_{S}  - g_{S,\text{B}}   )}{g_{S,\text{B}}}     G(P_{\text{B}},L)           + \frac{ g^2 G^{ 0}  (P_{\text{B}},L)}{ E_{\text{B}}}  + \mathcal O(e^{- \sqrt{2}  \kappa_{\text{B}} L}) \,.
\end{align}
The second term on the right hand side contains infinite volume quantities evaluated at the bound state pole mulitplied by the finite volume correction to the L\"uscher pole, while the third and fourth terms are corrections from the $L\to\infty$ expansion of the $G$ functions. 

As an illustration of the numerical effects, we consider the finite volume spectrum and matrix element for the proton-neutron scattering parameters, shown in Fig.~\ref{fig:fig-energies}. We use an effective range expansion to parameterize scattering amplitude, and assume that the single particle and bound state scalar charges are identical. We compare the all-orders expressions, Eqs.~\eqref{eq:luscher} and \eqref{eq:fv_matrix_simple_scalar}, to the leading order expressions obtain from Eqs. \eqref{eq:delta_sB} and \eqref{eq:BH_scalar_expand}. As seen in Fig.~\ref{fig:fig-energies}, there are significant deviations from the $L\to\infty$ expansion as compared to the all-orders expression, Eq.~\eqref{eq:fv_matrix_simple_scalar}. This illustrates a crucial need to consider general frameworks as in Refs.~\cite{Briceno:2015tza, Baroni:2018iau} in order to correctly asses the volume dependence.

\section{Summary}
In summary, the study presented provides two non-trivial checks on the recent formalism to extract infinite volume $\2+\Jc\to\2$ amplitudes from Lattice QCD~\cite{Briceno:2015tza, Baroni:2018iau}.  First, we observed that the charge of a conserved vector current remains independent of finite volume corrections, which is built into the formalism via the Ward-Takahashi identity. Second, we examined the $L\to\infty$ limit of a bound state scalar charge. Finite volume corrections arise from both the spectra and the $G$ functions of the formalism. For details, we refer the reader to Ref.~\cite{Briceno:2019nns}. More studies are underway investigating the near threshold expansion of finite volume matrix elements via this formalism, as well as a detailed study of the analytic behavior of $\2+\Jc\to\2$ transition amplitudes and resonance/ bound state form-factors.

\section{Acknowledgements}
I would like to thank my collaborators Ra\'ul A. Brice\~no and Maxwell T. Hansen for their contributions in completing this work.
I also thank Alessandro Baroni, Felipe Ortega-Gama, and Akaki Rusetsky for useful discussions.
This work is supported in part by USDOE grant No. DE-AC05-06OR23177, 
under which Jefferson Science Associates, LLC, manages and operates Jefferson Lab.
This work is also supported from the USDOE Early Career award, contract de-sc0019229. 

\bibliographystyle{JHEP}
\bibliography{bibi}
%



\end{document}